\documentclass[aps,12pt]{revtex4}

\begin{document}

\title{
Nuclear Three-body Force Effect on a Kaon Condensate in Neutron
Star Matter}

\author{W.~Zuo$^{1,2,3,4}$, A.~Li$^{2}$, Z.~H.~Li$^{1,3}$, U.Lombardo$^{5}$}

\affiliation{ $^1$ Institute of Modern Physics, Chinese Academy of
Sciences,
Lanzhou 730000, P.~R.~China\\
$^2$ School of Physics and Technology, Lanzhou University,
Lanzhou 730000, P.~R.~China\\
$^3$ Graduate School of Chinese Academy of Sciences,
Beijing 100039, P.~R.~China\\
$^4$ Institut f\"ur Theoretische Physik der Justus-Liebig-Universit\"at,
D-35392, Giessen, Germany\\
$^5$ INFN-LNS, Via Santa Sofia 44, I-95123 Catania, Italy}

\begin{abstract}

We explore the effects of a microscopic nuclear three-body force
on the threshold baryon density for kaon condensation in chemical
equilibrium neutron star matter and on the composition of the kaon
condensed phase in the framework of the Brueckner-Hartree-Fock
approach. Our results show that the nuclear three-body force
affects strongly the high-density behavior of nuclear symmetry
energy and consequently reduces considerably the critical density
for kaon condensation provided that the proton strangeness content
is not very large. The dependence of the threshold density on the
symmetry energy becomes weaker as the proton strangeness content
increases. The kaon condensed phase of neutron star matter turns
out to be proton-rich instead of neutron-rich. The three-body
force has an important influence on the composition of the kaon
condensed phase. Inclusion of the three-body force contribution in
the nuclear symmetry energy results in a significant reduction of
the proton and kaon fractions in the kaon condensed phase which is
more proton-rich in the case of no three-body force. Our results
are compared to other theoretical predictions by adopting
different models for the nuclear symmetry energy. The possible
implications of our results for the
neutron star structure are also briefly discussed.\\[3mm]
\end{abstract}
\maketitle

\section{Introduction}
As well known that around $\rho \simeq  \rho_0 $ where $\rho_0 =
0.16 ~{\rm fm}^{-3}$ is the empirical saturation density of
nuclear matter, neutron star matter in chemical equilibrium
consists mainly of nucleons and leptons, whereas at high enough
densities it may become more or less exotic\cite{pethick:1992}.
Among the possible exotic phases in dense nuclear matter, the kaon
condensation is a subject of great interest in nuclear physics,
hadronic physics and neutron star
physics\cite{lee:1996ef,li:1997,brown:1998}. Since been proposed
by Kaplan and Nelson\cite{kaplan:1986yq}, the possibility of kaon
condensation and its implications for astrophysical phenomena of
neutron stars have been extensively discussed by many
physicists\cite{politzer:1991,brown:1992,glendenning,
thorsson:1994,ellis:1995,yasuhira:2000,
pons:2001,ramos:2001,carlson:2001,norson:2001,kubis:2003}. It has
been suggested in Ref.\cite{bkr} that the condensation of kaons
may be understood as a chiral rotation away from a V-spin scalar
ground state and it is related to a partial restoration of the
chiral symmetry explicitly broken in the vacuum. The presence of
kaon condensation may have important consequences for determining
the structures and evolutions of neutron stars. For example, the
kaon condensation may soften substantially the equation of state
(EOS) of neutron star matter and consequently lower the predicted
maximum mass of neutron stars\cite{li:1997}. The phase transition
from the normal matter to the kaon condensed matter is also
expected to affect the transport properties and the glitch
behavior of pulsars\cite{glendenning}.

It is shown \cite{kaplan:1986yq} that the kaon-nucleon sigma term
$\Sigma_{KN}$ provides a strongly attractive interaction between
kaons and nucleons. This attraction reduces the $K^-$ energy
$\omega_K$ in nuclear medium and it becomes strong enough at a
high enough baryon density above which the kaon condensed phase is
energetically favorable. It is shown that the kaon--nucleon sigma
term plays an essential role in determining the formation of such
a phase. Besides the kaon-nucleon interaction, the high-density
behavior of the nuclear symmetry energy is also important for
determining the threshold density for kaon condensation and the
composition of the kaon condensed phase. Associated to the
high-density symmetry energy there is a large uncertainty due to
the lack of experimental constraints\cite{danielewicz}. The
nuclear symmetry energy plays its role somewhat in a different way
from the kaon--nucleon interaction. It determines to a large
extent the electron chemical potential in $\beta$-equilibrium
neutron star matter below the critical density for
kaon-condensation\cite{lattimer:1991,bombaci:1991,lejune:2000}.
The electron chemical potential $\mu_e$ serves as the energy
threshold for kaon condensation since the kaon condensation
becomes energetically favorable as soon as $\mu_e > \omega_K$.

Since realistic nucleon-nucleon (NN) interactions have not been
fully determined from the chiral theory up to
now\cite{kaiser:2002}, various theoretical models of the symmetry
energy have been adopted for studying the properties of kaon
condensation. In the work of Ref.\cite{thorsson:1994}, the simple
parametrizations proposed by Prakash et al.\cite{prakash:1988} has
been chosen for the nuclear symmetry energy. In the investigations
of Ref.\cite{kubis:2003}, the symmetry energy derived from the
variational approach\cite{wiringa:1988tp} has been applied. It
turns out that besides the dependence on $\Sigma_{K N}$, the
condensation is also sensitive to the high--density behavior of
the symmetry energy. Within the framework of a non-relativistic
microscopic model based on realistic NN interactions, nuclear
three-body forces are critical for reproducing the empirical
saturation properties of nuclear
matter\cite{baldo:1997,grange:1989,machleidt:1989,wei:2002,fuchs:2003}.
In our previous work\cite{lejune:2000,wei:2002}, the EOS of
nuclear matter has been explored in the Brueckner-Hartree-Fock
(BHF) approach by adopting a microscopic three-body force (TBF)
from the meson-exchange current method~\cite{grange:1989}. The TBF
turns out to affect strongly the high-density behavior of the
symmetry energy, i.e., it makes the density dependence of the
symmetry energy much stiffer as compared to the result without the
TBF contribution. The aim of the present work is to investigate
the critical density for kaon condensation in neutron stars and
the composition of the kaon condensed phase in neutron star matter
by using the BHF approach for the nuclear symmetry energy. In the
calculations, we adopt the chiral Lagrangian\cite{kaplan:1986yq}
to extract the kaon-nucleon part of the interactions as in
Refs.\cite{thorsson:1994,kubis:2003}. Special attention has been
paid on the effects of the nuclear three-body force.

This paper is organized as follows. In Sect.~2 we review briefly
the theoretical models adopted in our calculations, including the
self-consistent BHF approach, the microscopic TBF, and the chiral
model for kaon-nucleon interaction. Our numerical results are
presented and discussed in Sect.~3. In Sect.~4 a summary of the
present work is given.

\section{Theoretical models}\label{sec_symmen}

In the kaon condensed phase of neutron star matter,
the energy density consists of three parts of contributions, i.e.
\begin{equation}
\varepsilon = \varepsilon_{NN} + \varepsilon_{lep} +
\varepsilon_{KN}
\end{equation}
where $\varepsilon_{NN}$ is the
nuclear part, $\varepsilon_{lep}$ denotes the contribution of
leptons, and $\epsilon_{KN}$ is the contribution from the
kaon-nucleon interaction.

Our calculations of the nuclear part of the EOS are based on the
BHF approach for asymmetric nuclear matter \cite{bombaci:1991}.
The starting point of the BHF approach is the interaction $G$
matrix which satisfies the following Bethe-Goldstone (BG)
equation\cite{bombaci:1991}
\begin{equation} G(\rho,\beta;\omega)
= v_{NN} + v_{NN} \sum_{k_1 k_2}
 \frac { |k_1 k_2 \rangle Q(k_1,k_2) \langle k_1 k_2| }
{\omega - \epsilon(k_1)-\epsilon(k_2)+i\eta} G(\rho,\beta;\omega),
\label{eq:BG}
\end{equation}
where $\omega$ is the starting energy and $Q(k_1,k_2) = [1-n(k_1)]
[1-n(k_2)]$ is the Pauli operator which prevents the two
intermediate nucleons from being scattered into the states below
the Fermi sea. The isospin asymmetry parameter is defined as
$\beta=(\rho_n-\rho_p)/\rho_B$, $\rho_n$, $\rho_p$, and $\rho_B$
being the neutron, proton and total baryon number densities,
respectively. The single particle (s.p.) energy is given by
$\epsilon(k) = \hbar^2k^2/(2m)  +  U(k)$, where the s.p. potential
is calculated from the real part of the on-shell G matrix, i.e.
\begin{equation} U(k) = \sum_{k'} n(k') {\rm Re} \langle k k'|G(\epsilon(k)
+ \epsilon(k')) |k k'\rangle_A \ . \label{eq:U1}
\end{equation}
In the present calculations, we adopt the continuous choice for
$U(k)$ since it has been proved to provide a much faster
convergency of the hole-line expansion for the energy per nucleon
in nuclear matter up to high densities than the gap
choice~\cite{song:1998}. In addition, in the continuous choice,
the s.p. potential describes physically the nuclear mean field
felt by a nucleon in nuclear medium. The realistic nucleon-nucleon
(NN) interaction $v_{NN}$ is the Argonne $V_{18}$ ($AV_{18}$)
two-body force\cite{wiringa:1995} supplemented with the
contribution of the TBF.

The microscopic TBF adopted in the present calculations is
constructed from the meson-exchange current
approach~\cite{grange:1989}. In this TBF model, four important
mesons $ \pi $, $ \rho $, $ \sigma $ and $\omega $ are
considered\cite{machleidt:1989}. The meson masses in the TBF have
been fixed at their physical values except for the virtual
$\sigma$-meson mass which has been fixed at $540$MeV according to
Ref.~\cite{grange:1989}. This value has been checked to
satisfactorily reproduce the $AV_{18}$ interaction from the
one-boson-exchange potential (OBEP) model\cite{wei:2002}. The
other parameters of the TBF, i.e., the coupling constants and the
form factors, have been determined from the OBEP model to meet the
self-consistent requirement with the adopted $AV_{18}$ two-body
force. The values of the parameters are given in
Ref.\cite{wei:2002}. For a more detailed description of the model
we refer to Refs.~\cite{grange:1989,wei:2002}. We want to stress
that the most important component of this TBF is essentially that
introduced automatically in the Walecka relativistic mean field
theory~\cite{walecka:1974} over that in the non relativistic
theories.

In order to include the TBF contribution into the BG equation, we
follow the standard
scheme\cite{grange:1989,lejeune:1986,baldo:1998} to reduce the TBF
to an effective two-body interaction which is expressed in $r$
space as
\small \begin{equation}
\begin{array}{lll}
 \langle\vec r_1 \vec r_2| V_3^{\tau_1\tau_2} |
\vec r_1^{\ \prime} \vec r_2^{\ \prime} \rangle &=&
\displaystyle
 \frac{1}{4}
{\rm Tr}\sum_n \int {\rm d}
{\vec r_3} {\rm d} {\vec r_3^{\ \prime}}\phi^*_n(\tau_3\vec r_3^{\
\prime}) (1-\eta_{\tau_1,\tau_3}(r_{13}' ))
(1-\eta_{\tau_2,\tau_3}(r_{23}')) \\[6mm]
&\times& \displaystyle W_3(\vec r_1^{\ \prime}\vec r_2^{\ \prime}
\vec r_3^{\ \prime}|\vec r_1 \vec r_2 \vec r_3) \phi_n(\tau_3r_3)
(1-\eta_{\tau_1,\tau_3}(r_{13}))(1-\eta_{\tau_2,\tau_3}(r_{23}))
\end{array}\label{eq:TBF}
\end{equation} \normalsize
where the trace is taken with respect to the
spin and isospin of the third nucleon and the indices $\tau_1,
\tau_2$ and $\tau_3$ denote the isospin $z$-components.. The
function $\eta_{\tau,\tau\prime}(r)$ is the defect function. Since
the defect function is directly determined by the solution of the
BG equation\cite{grange:1989,lejeune:1986}, it must be calculated
self-consistently with the $G$ matrix and the s.p. potential
$U(k)$\cite{wei:2002} at each density and isospin asymmetry.

By solving self-consistently the coupled Eqs.(\ref{eq:BG}),
(\ref{eq:U1}), and (\ref{eq:TBF}) we can obtain the reaction $G$
matrix. From the $G$ matrix, we calculate the nuclear contribution
$\varepsilon_{NN}$ to the total energy density $\varepsilon$ for
any given baryon density and isospin asymmetry. The
$\varepsilon_{NN}$ of asymmetric nuclear matter can be separated
into two part, i.e., the isoscalar part and the isovector part.
The stiffness of the isoscalar part plays an important role in
predicting the structure of a neutron star while the isovector
part is crucial for the chemical properties of neutron star
matter. Microscopic investigations\cite{bombaci:1991,wei:2002}
show that for a fixed baryon density and isospin asymmetry, the
isovector part of $\varepsilon_{NN}$ is completely determined by
the symmetry energy $S(u)$ and can be expressed as $\rho_0 u
S(u)\beta^2$, where $u\equiv\rho_B/\rho_0$ is the dimensionless
baryon density and $\rho_0=0.16$ fm$^{-3}$ the empirical
saturation density of nuclear matter.

We evaluate the kaon-nucleon sector in the energy density of
the kaon-condensed phase of matter in terms of the effective chiral
Lagrange density which was suggested by Kaplan and
Nelson~\cite{kaplan:1986yq} and extensively investigated
afterwards\cite{brown:1992}. In the Kaplan-Nelson model, the
$SU(3)\times SU(3)$ chiral Lagrange density is expressed as
\begin{eqnarray}
{\mathcal  L_\chi} & = & \frac{f^2}{4}{\mathrm Tr} \partial_\mu U
\partial^\mu
U^\dagger +{\mathrm Tr}\bar{B}(i \gamma^\mu D_\mu -  m_B) B  \nonumber \\
& & + \; F\, {\mathrm Tr} \bar{B}\gamma^\mu \gamma_5 [ {\mathcal
A}_\mu,B] + \; D\, {\mathrm Tr} \bar{B}\gamma^\mu \gamma_5 \{
{\mathcal A}_\mu,B\}
\label{lan-KN} \nonumber \\
& & + c {\mathrm Tr} {\mathcal M} (U + U^\dagger) + a_1 {\mathrm
Tr} \bar{B}(\xi {\mathcal M} \xi + \xi^\dagger {\mathcal M}
\xi^\dagger) B \nonumber \\
& & + a_2 {\mathrm Tr} \bar{B}B(\xi {\mathcal M} \xi + \xi^\dagger
{\mathcal M} \xi^\dagger) + a_3 {\mathrm Tr} \bar{B}B\; {\mathrm
Tr}(\xi {\mathcal M} \xi + \xi^\dagger {\mathcal M} \xi^\dagger).
\end{eqnarray}
where we adopt the same notations and symbols as in
Ref.\cite{kaplan:1986yq}. The above Lagrange includes an octet of
pseudoscalar masons $\mathcal M$ and an octet of baryons $B$. The
first four terms conserve the chiral symmetry. $f=93\mathrm{MeV}$
is the pion decay constant. The constants $D$ and $F$ are given by
$D=0.81$ and $F=0.44$. The coefficients $a_1m_s, a_2m_s$,
$a_3m_s$, and $c$ determine the strength of the chiral symmetry
breaking in the last four terms. We adopt $a_1m_s=-67$ MeV and
$a_2m_s=134$ MeV obtained from the baryon mass
splittings\cite{politzer:1991}. The constant $c$ is related to the
bare kaon mass by the Gell-Mann-Oakes-Renner relation $m_K^2 =
2cm_s/f^2$. The parameter $a_3m_s$ had remained largely uncertain
for many years due to our poor knowledge of the strangeness
content of the proton and the kaon-nucleon sigma term
$\Sigma_{KN}$\cite{lee:1996ef,kaplan:1986yq,politzer:1991,donoghue:1985bu}.
Fortunately, the large ambiguity in this parameter has been
settled recently by Dong, Laga\"{e} and Liu~\cite{dong:1996} with
small error based on the lattice calculations. In the present
calculations we adopt the value of $a_3m_s$ in the range $-310 \
\mathrm{MeV} < a_3 m_s < -134  \ \mathrm{MeV}$ as done
 in Refs.\cite{thorsson:1994,kubis:2003} in order to investigate the
 sensitivity of the TBF effect to the
 varying of the proton strangeness content. Our adopted
central value $a_3m_s =-222$ MeV is very close to the value
extracted from the lattice guage calculations in
Ref.~\cite{dong:1996} which is $-231$ MeV with error of less than
$4\%$. $a_3m_s$ measures the proton
 strangeness content and plays an essential role in determining
 the threshold density for kaon condensation\cite{thorsson:1994} since
 it provides the dominant attraction in the kaon-nucleon interaction via the
 sigma term $\Sigma_{KN} = -(a_1+2a_2+4a_3)m_s/2$. The above specified range
 of $a_3m_3$ from $-310 \ \mathrm{MeV}$ to $ -134 \ \mathrm{MeV}$
 corresponds to a reasonable range $0.2>\langle\bar s s\rangle_p/
 \langle\bar u u + \bar d d + \bar s s\rangle_p>0$ of
 the proton strangeness content\cite{donoghue:1985bu}.
Following
exactly the standard prescript in Ref.\cite{thorsson:1994},
we can get the kaon-nucleon energy density of the kaon condensed matter, i.e.
\begin{equation}
\varepsilon_{KN} = \frac{f^2}{2} \mu_K^2\sin^2\theta
+ 2m^2_Kf^2\sin^2\frac{\theta}{2}+
\rho_B(2a_1x_p+2a_2+4a_3)m_s\sin^2\frac{\theta}{2}
\end{equation}
where $\rho_B$ and $x_p\equiv\rho_p/\rho_B$ denote the baryon number density
and the proton fraction, respectively.
$\theta$ is the amplitude of the condensation and comes from applying the Baym
theorem\cite{baym}.
The lepton part of the energy density $\epsilon_{lep}$
can be readily obtained in a
standard way from the contributions of the filled Fermi seas of
leptons\cite{thorsson:1994,kubis:2003}.

In the neutron star matter with kaons the chemical equilibrium can
be reached through the following reactions
\begin{equation} n
\leftrightarrow p + l + \nu_l, ~~~ n \leftrightarrow  p + K^-
,\label{beta} ~~~ l \leftrightarrow  K^- + \nu_l ,\nonumber
\end{equation}
where $l$ denotes leptons, i.e., $l = e,\mu$. One can determine
the ground state by minimizing the total energy density with
respect to the condensate amplitude $\theta$ keeping all densities
fixed. This minimization together with the chemical equilibrium
and charge neutrality conditions leads to the following three
coupled equations\cite{thorsson:1994,kubis:2003}
\begin{equation}
\cos\theta = \frac{1}{f^2 \mu^2} \left(m_K^2 f^2 +
\frac{1}{2}u\rho_0(2a_1 x\!+\!2a_2 \!+\!4a_3)m_s
    - \frac{1}{2}\mu u\rho_0 (1\!+\!x) \right)~,
\label{th-min}
\end{equation}
and
\begin{equation} \mu \equiv
\mu_e=\mu_K = \mu_n-\mu_p= 4(1-2x)S(u)\sec^2 \frac{\theta}{2} -
2a_1 m_s \tan^2\frac{\theta}{2} \,, \label{betatheta}
\end{equation} \begin{equation}
 f^2\mu
 \sin^2\theta + u\rho_0 (1+x)\sin^2\frac{\theta}{2} - xu\rho_0
 + \frac{ \mu^3 }{ 3\pi^2 }
 + \eta(|\mu|-m_\mu) \frac{ (\mu^2-m_\mu^2)^{3/2}}{ 3\pi^2 } = 0,
\nonumber \\
\label{cntheta} \end{equation}
where $\mu$ represents the chemical
potential of the negative charged particles and
$u\equiv\rho_B/\rho_0$ is the baryon number density in unit of
$\rho_0$. The last two equations are from the chemical equilibrium
and charge neutrality conditions, respectively. The EOS and the
composition of the kaon condensed phase in the chemical
equilibrium neutron star matter can be obtained by solving the
coupled equations (\ref{th-min}),
 (\ref{betatheta}), and (\ref{cntheta}). The critical density for
kaon condensation is determined as the very point above which a
real solution for the coupled equations can be found.

\section{Results and Discussions}
\label{sec_thermod}

The density dependence of the nuclear symmetry energy $S(u)$ is
depicted in Fig.\ref{fig1} where the bold solid line and the thin
solid line are obtained from the BHF approach adopting the
$AV_{18}$ plus the TBF
 and the pure $AV_{18}$ two-body force, respectively.
In the figure the symmetry energy from other
 models\cite{prakash:1988,wiringa:1988tp}
are also plotted for comparison. The dashed, dotted, and
dot-dashed curves correspond to the following parameterizations
proposed in Ref.\cite{prakash:1988},
\begin{equation} S(u) =
\left( 2^{\frac{2}{3}}-1 \right) \frac{3}{5}E_F^{(0)}\left(
u^{\frac{2}{3}} - F(u) \right)
     + S_0 F(u)
\end{equation}
and
\begin{equation} F_1(u) = u, \qquad F_2(u) = \frac{2u^2}{1+u}
\qquad {\rm or} \qquad F_3(u) = \sqrt{u} \label{fu}
\end{equation}
where $S_0 \simeq 30$ MeV and $E_F^{(0)} =
(3\pi^2\rho_0/2)^{2/3}/2m$ are the symmetry energy and the Fermi
energy at the nuclear saturation density, respectively. The
double-dot-dashed line is the result of Ref.\cite{wiringa:1988tp}
by using the variational
 approach with the UV14+TNI interaction.
It is seen from the figure that in both cases with and without the
TBF contribution, our calculated $S(u)$ are monotonically
increasing functions of the baryon density. Below and around the
empirical saturation density, the TBF effect is quite small, while
at high densities the TBF contribution leads to a strong
enhancement of the increasing rate of $S(u)$. This strong
enhancement of the symmetry energy due to the TBF could be
explained as follows. The symmetry energy (the isovector sector of
the EOS) describes the energy required to increase the isospin
asymmetry of the matter. A higher value of asymmetry implies a
larger neutron excess in the matter. Therefore for a fixed
density, at a higher isospin asymmetry the neutron Fermi momentum
becomes larger and consequently some nucleons may have higher
momentum than in the symmetric nuclear matter. Since the TBF is a
short-range interaction and its effect is stronger for nucleons
with larger momenta, its contribution to the EOS increases as the
isospin asymmetry increases~\cite{grange:1989,wei:2002}.

The density dependence of the chemical potential $\mu$ of the
negative charged particles in $\beta$-equilibrium neutron star
matter is given in Fig.\ref{fig2} for three values of
$a_3m_s=-310, -222$, and $-134$ MeV denoted by $a, b, c$,
respectively. In the figure the solid and dashed curves are the
results with and without including the nuclear TBF contribution.
Below the critical density for kaon condensation, the matter is
made up of neutrons, protons, and leptons. The chemical potential
$\mu$ is determined by the density dependence of the symmetry
energy and it increases gradually with density in the case that
the symmetry energy is an monotonically increasing function of
density. The TBF enhances the increasing rate of $\mu$. Above the
critical density, $\mu$ becomes a decreasing function of density.
The decreasing rate of $\mu$ depends strongly on the choice of the
proton strangeness content (i.e., on the value of $a_3m_s$), but
completely insensitive to the choice of different models for the
symmetry energy $S(u)$ in the case that $S(u)$ is an monotonically
increasing function of density as shown in
Ref.\cite{thorsson:1994}. As a consequence, the TBF contribution
has almost no any effect on the chemical potential in the
condensed phase.

The values of the critical density $u_c$ for different models of
the symmetry energy $S(u)$, and some typical values of the proton
strangeness are presented in Tab.\ref{tab-nc}. The results in
Ref.\cite{thorsson:1994} (denoted with $F1, F2$, and $F3$) and
Ref.\cite{kubis:2003} (denoted with $UV14+TNI$) are also given.
\begin{table}[ht]
\begin{center}
\caption{\small Critical density $u_c$ for kaon condensation in
unit of $\rho_0$. } \label{tab-nc}
\end{center}
\begin{center}
\begin{tabular}{|c|c|c|c|c|c|c|c|}
\hline
   &  $~a_3 m_s~[\mathrm{MeV}]$
   &  F1 & F2 & F3 & BHF & BHF+TBF & UV14+TNI \\
\hline
a & $-310$ & 2.4 & 2.3 & 2.6 & 2.6 & 2.4 & 2.8 \\
b & $-222$ & 3.1 & 2.9 & 3.4 & 3.4 & 2.9 & 4.1 \\
c & $-134$ & 4.2 & 3.8 & 4.9 & 5.0 & 3.8 &  -- \\
\hline
\end{tabular}
\end{center}
\end{table}
As shown in Tab.\ref{tab-nc}, the critical density $u_c$ for kaon
condensation depends on both the proton strangeness content (i.e.,
the value of $a_3m_s$) and the high-density behavior of the
symmetry energy. The critical density $u_c$ is more sensitive to
the proton strangeness than to the symmetry energy. The value of
$u_c$ turns out to be lowered by increasing the proton
strangeness, since the attractive interaction provided by the
kaon-nucleon sigma term becomes smaller as the strangeness
increases. This is in good agreement with the previous
investigations by adopting different models for the symmetry
energy\cite{kaplan:1986yq,thorsson:1994}. It is seen from the
table that the predicted values of $u_c$ become less sensitive to
different models for the symmetry energy as the strangeness
increases. The TBF affects the critical density $u_c$ via its
contribution to the symmetry energy and its effect is to reduce
the critical density since it provides an additional repulsive
contribution to the isospin symmetry energy\cite{wei:2002}. Our
predicted critical density is in the range of $2.4\rho_0 - 3.8
\rho_0$ if the TBF is included and $2.6 \rho_0 - 5.0\rho_0$ in the
case without the TBF contribution. In the case of no strangeness,
inclusion of the TBF contribution in the symmetry energy reduces
$u_c$ by more than $20\%$ from 5 to 3.8. However, if the
strangeness is higher, the TBF effect on $u_c$ becomes somewhat
smaller. For instance, in the case of $a_3m_s=-310$MeV, the TBF
leads to only a less than $10\%$ reduction of $u_c$. This can be
readily understood since on the one hand, for larger values of the
strangeness, the kaon condensation sets in at lower densities
where the TBF contribution to the symmetry energy is relatively
small. On the other hand, if the strangeness is higher, the
in-medium energy of $K^-$ drops down faster as a function of
baryon density. As a consequence, the role played by the
strangeness becomes more predominant over that by the symmetry
energy for a higher value of the strangeness.

In Fig.\ref{fig3} is plotted the predicted composition of
$\beta$-equilibrium neutron star matter for three different values
of $a_3m_3=-310$ MeV(the upper panel), $-222$ MeV(the middle
panel), and $-134$ MeV(the lower panel) for both cases with the
TBF (curves with symbols) and without the TBF (curves without
symbols). In the figure the proton fraction
$x_{p}\equiv\rho_p/\rho_B$, the kaon fraction
$x_{K}\equiv\rho_K/\rho_B$, and the lepton fraction
$x_{lep}=x_e+x_{\mu}\equiv (\rho_e+\rho_{\mu})/\rho_B$ are given
by the solid curves, dashed curves, and dotted curves,
respectively. Below the critical density, the matter is in its
normal phase which is a highly neutron-rich and charge-neutral
mixture of neutrons (n), protons (p), electrons (e), and muons
($\mu$). In the normal $n, p, e, \mu$ matter, the proton fraction
$x_p$ increases gradually with density and the TBF contribution
makes the $x_p$ rise faster. However, in the condensed phase, the
matter becomes more or less proton-rich instead of neutron-rich,
since a large fraction of protons is required to balance the
negative charge of $K^-$ which is so abundant in the kaon
condensed phase. At high enough density, the matter contains even
positrons in chemical equilibrium to ensure the charge neutrality.
These results are in agreement with the previous
investigations\cite{kubis:2003,thorsson:1994}. From
Fig.\ref{fig3}, one can see that the proton and kaon fractions in
the condensed phase depend sensitively on the high-density
behavior of the symmetry energy. Inclusion of the TBF contribution
makes the kaon condensed neutron star matter more symmetric in
protons and neutrons. As the matter goes from the normal phase to
the kaon condensed phase, the redistribution of charge depends
strongly on the high-density behavior of the symmetry energy. If
the TBF contribution is included, the symmetry energy rises much
faster as a function of density. As a consequence, the matter is
driven to more symmetric in neutrons and protons in order to
reduce the additional repulsive isospin energy due to the TBF
contribution.

In Ref.\cite{kubis:2003} the kaon condensation has been studied by
adopting the symmetry energy from the variational calculations in
Ref.\cite{wiringa:1988tp}. In contrast to the present results from
the BHF approach as well the predictions from the Dirac-Brueckner
method\cite{huber:1998} and the relativistic mean field
theory\cite{greco:2001}, the symmetry energy calculated in
Ref.\cite{wiringa:1988tp} decreases with density at high enough
densities. As a consequence, the composition in the kaon condensed
phase in Ref.\cite{kubis:2003} strongly deviates from our result
due to the completely different high-density behavior of the
symmetry energy. For example, in the most extreme case of the
UV14+TNI interaction, the obtained neutron star becomes almost a
``proton" star just above the critical density for kaon
condensation as shown in Fig.~4. While in the present
calculations, the neutron star matter with kaons prefers to be
symmetric nuclear matter in agreement with the results reported in
Ref.\cite{thorsson:1994}. The discrepancy between the high-density
behavior of the symmetry energy obtained from the BHF and the
variational approaches is still not
clear\cite{bombaci:1991,baldo:1997} and deserves further
investigations.

It should be stressed that in the present investigations, we do not consider
the mixed phase of the normal matter and the condensed matter. It has been
pointed out\cite{glendenning} that in the first-order phase
transition to the kaon condensed phase, the conservation law(s) can only be
fulfilled in a global rather than a local sense. This means that in the
mixed phase, the local charge-neutrality, i.e. Eq.(\ref{cntheta}) becomes invalid.
Since our aim is to investigate the nuclear TBF effects on the critical
density and the chemical composition of the condensed phase,
the present simplification is somewhat desirable for our purpose. However,
for practical applications in neutron stars we should consider the problem
of the mixed phase formation as done in Ref.\cite{glendenning}
which will be discussed elsewhere.

Before summary, we discuss briefly the possible implications of
our results for the maximum mass of neutron stars. In
Fig.\ref{fig5} we display the predicted EOS of neutron star matter
for both cases with kaons (solid curves) and without kaons (dashed
curves). In the figure the results by using the $AV_{18}$ plus the
TBF (right panel) are compared to the ones by adopting the pure
$AV_{18}$ two-body force (left panel). It is seen that in the case
of no kaons, inclusion of the TBF contribution makes the EOS at
high densities become much more stiffer. As a consequence, the
maximum mass of neutron stars becomes much larger if the TBF
contribution is included as shown in Ref.\cite{zhou:2004} where
the calculated maximum mass is about 2.3 M$_\odot$ and 1.6
M$_\odot$
 respectively for the two cases
with and without including the TBF in the calculations for our
central value of $a_3m_s = -222$ MeV. The latter value is close to
the 1.5 M$_{\odot}$ found in Ref.\cite{thorsson:1994} for a
compression modulus of 210 MeV. In the case that kaons are
allowed, the EOS of the matter in the condensed phase is softened
considerably due to the attractive interaction between nucleons
and kaons. Since a softer EOS implies a smaller maximum neutron
star mass, the kaon condensation may lower the predicted maximum
mass in both cases with and without the TBF. In our present work,
however, we do not consider the role of other possible strange
particles, such as $\Lambda$, $\Sigma$, $\Xi$ hyperons. The
appearance of hyperons may provide an additional softening of the
neutron star EOS and consequently reduce further the predicted
maximum mass\cite{baldo:1998prc}. One would have to minimize the
free energy at each density in order to choose the appropriate
configuration.

\section{Summary and Conclusion}
In summary, we have investigated the nuclear TBF effects on the
kaon condensation in neutron star matter in the framework of the
BHF approach. The TBF affects the critical density for the kaon
condensation through its repulsive contribution to the symmetry
energy. In both cases with and without the TBF, the calculated
symmetry energy is an monotonically increasing function of baryon
density, in agreement with the results from the relativistic mean
field approach and the DB approach. The TBF repulsion turns out to
increases rapidly as the baryon density increases. As a
consequence the high-density behavior of the symmetry energy
becomes much stiffer as compared to the one obtained by adopting
only the two-body nuclear force. Our results show that inclusion
of the TBF contribution in the symmetry energy reduces the
critical density by about $8\%\sim25\%$ depending on the choice of
the proton strangeness content. The influence of the TBF on the
critical density becomes smaller if the strangeness is higher. The
predicted critical density is in the range from 2.4$\rho_0$ to
3.8$\rho_0$ when the TBF is included.

The additional repulsive contribution to the isospin energy due to
the TBF drives the kaon condensed phase of neutron star matter to
become more symmetric in neutrons and protons as compared to the
results without the TBF. In the normal phase of neutron star
matter the proton fraction is small and the matter is highly
neutron-rich. The proton fraction gradually increases with baryon
density. While in the kaon condensed phase, the matter becomes
proton-rich in order to balance the negative charge of the kaon
field. The composition in the kaon condensed phase depends
sensitively on the high-density behavior of the symmetry energy.
We find that the additional repulsion from the TBF lowers the
proton and kaon fractions in the kaon condensed phase and it is so
strong at high densities as to make the condensed matter almost
symmetric in neutrons and protons. The EOS of neuron star matter
is found to be softened considerably by the kaon-nucleon
interaction in the kaon condensed phase.


\section{Acknowledgment}

The work is supported in part by the Knowledge Innovation Project
of the Chinese Academy of Sciences (KJCX2-SW-N02), the Major
State Basic Research Development Program of
 China (G2000077400), and the Major Prophase Research Project of
Fundamental Research of the Ministry of Science and Technology of
China (2002CCB00200), the National Natural Science Foundation of
China (10235030, 10175082) and DFG, Germany.


\begin{figure}
\caption{\small The density dependence of nuclear symmetry energy
from different models: F1(dashed curve), F2(dotted curve),
F3(dot-dashed curve), UV14+TNI(double-dot-dashed line), BHF(thin
solid line), and BHF+TBF(bold solid line). \label{fig1} }

\caption{\small The chemical potential of the negative charged
particles for three different values of $a_3m_s=-310, -222$, and
$-134$ MeV denoted by a,b and c respectively. The solid curves
represent the results including the TBF contribution, while the
dashed curves are the results without the TBF contribution.
\label{fig2}}

\caption{\small The composition of chemical equilibrium neutron
star matter for the two cases with (curves with symbols) and
without (curves without symbols) the TBF contribution. The solid
curves are the results for the proton fraction $x_p$, the dashed
curves for kaon fraction $x_K$, and the dotted curves for lepton
fraction $x_{l}$.  \label{fig3}}

\caption{The composition of neutron star matter by using different
models for the symmetry energy and adopting $a_3m_s=-222$ MeV. The
curves with symbols are obtained from the BHF calculations
including the TBF contribution (BHF+TBF). The bold curves without
symbols correspond to the results of Ref.\cite{kubis:2003} by
adopting the symmetry energy of the variational approach and the
UV14+TNI interaction\cite{wiringa:1988tp}. The thin curves are
obtained by using the linear density dependent symmetry
energy\cite{thorsson:1994}. The solid and dashed curves are the
results for the proton fraction $x_p$, and the lepton fraction
$x_{l}$ respectively. \label{fig4}}

\caption{The predicted EOS of neutron star matter for both cases
with kaons (solid curves) and without kaons (dashed curves). The
results by using the $AV_{18}$ plus the TBF (right panel) are
compared to the ones by adopting the pure $AV_{18}$ two-body force
(left panel). \label{fig5}}
\end{figure}

\begin{thebibliography}{13}

\bibitem{pethick:1992}C.~J.~Pethick, Rev.~Mod.~Phys.
{\bf 64}, 1133 (1992).

\bibitem{lee:1996ef} C.~H.~Lee, Phys. Rep. {\bf 275}, 255 (1996)
 and references therein; C. H. Lee, G. E. Brown, D. P. Min, and M. Rho,
 Nucl. Phys. {\bf A 585}, 401 (1995).

\bibitem{li:1997} G.Q.Li, C.-H.Lee, and G.E.Brown,
 Phys. Rev. Lett. {\bf 79}, 5214(1997); Nucl. Phys. {\bf A 625}, 372 (1997).

\bibitem{brown:1998} G. E. Brown, C.-H.Lee, and R. Rapp,
 Nucl. Phys. {\bf A 639}, 455c (1998).

\bibitem{kaplan:1986yq} D.~B.~Kaplan and A.~E.~Nelson,
 Phys. Lett. {\bf B 175}, 57 (1986).

\bibitem{politzer:1991} H. D. Politzer and M. B. Wise,
 Phys. Lett. {\bf B 273}, 156 (1991).

\bibitem{brown:1992} G.~E.~Brown, K.~Kubodera, M.~Rho, and V.~Thorsson,
 Phys. Lett. {\bf B 291}, 355 (1992).

\bibitem{glendenning} N. K. Glendenning, Phys. Rev. {\bf D 46}, 1274(1992);
 N. K. Glendenning and J. Schaffner-Bielich,
 Phys. Rev. Lett. {\bf 81}, 4564 (1998);
 Phys. Rev. {\bf C 60}, 025803 (1999).

\bibitem{thorsson:1994} V. Thorsson, M. Prakash, and J.~M.~Lattimer,
 Nucl. Phys. {\bf A 572}, 693(1994); {\bf A 574}, 851(1994), Erratum.

\bibitem{ellis:1995} P. J. Ellis, R. Knorren, and M. Prakash,
 Phys. Lett. {\bf B 349}, 11 (1995).

\bibitem{yasuhira:2000} M. Yasuhira and T. Tatsumi,
 Nucl. Phys. {\bf A 663}, 881(2000); {\bf A 670}, 218(2000);
 T. Muto, M. Yasuhira, T. Tatsumi, and N. Iwamoto,
 Phys. Rev. {\bf D 67}, 103002(2003).

\bibitem{pons:2001} J. A. Pons, J. A. Miralles, M. Prakash, and J. M. Lattimer,
 Astrophys. J. {\bf 553}, 382 (2001).

\bibitem{ramos:2001} A. Ramos, J. S. Bielich, and J. Wambach,
 Lett. Notes. Phys. {\bf 578}, 175 (2001).

\bibitem{carlson:2001} J. Carlson, H. Heiselberg, and V. R. Pandharipande,
 Phys. Rev. {\bf C63}, 017603 (2001).

\bibitem{norson:2001} T. Norsen and S. Reddy, Phys. Rev. {\bf C63}, 065804(2001).

\bibitem{kubis:2003} S.~Kubis and M.~Kutschera, Nucl. Phys. {\bf A 720}, 189(2003).

\bibitem{bkr} G.~E. Brown, K.~Kubodera, and M.~Rho,
 Phys. Lett. {\bf B 192}, 273(1987).

\bibitem{danielewicz} P. Danielewicz, R. Lacey, and W. G. Lynch,
 Science {\bf 298}, (2002)1592; B.~A.~Li, Phys. Rev. Lett.
 {\bf 88}, 192701 (2002); A. E. L. Dieperink, Y. Dewulf, D. Van Neck,
 M. Waroquier, and V. Rodin, Phys. Rev. {\bf C 68}, 064307 (2003).

\bibitem{lattimer:1991}J.~M. Lattimer, C. J. Pethick,
 M.~Prakash, and P. Haensel, Phys. Rev. Lett. {\bf 66}, 2701 (1991).

\bibitem{bombaci:1991} I.~Bombaci and U.~Lombardo, Phys. Rev. {\bf C 44}, 1892 (1991);
 W.~Zuo, I.~Bombaci, and U.~Lombardo,
 Phys. Rev. {\bf C 60}, 024605(1999).

\bibitem{lejune:2000} A. Lejeune, U. Lombardo, and W. Zuo, Phys. Lett. {\bf B477}, 45 (2000).

\bibitem{kaiser:2002} N. Kaiser, S. Fritsch, and W. Weise,
 Nucl. Phys. {\bf A 697}, 255 (2002).

\bibitem{prakash:1988} M.~Prakash, T.~L. Ainsworth, and J.~M. Lattimer,
Phys. Rev. Lett. {\bf 61}, 2518(1988).

\bibitem{wiringa:1988tp} R.~B.~Wiringa, V.~Fiks and A.~Fabrocini,
Phys. Rev. {\bf C 38}, 1010(1988).

\bibitem{song:1998} H.~Q.~Song, M.~Baldo, G.~Giansiracusa, and U.~Lombardo,
 Phys. Rev. Lett. {\bf 81}, 1584(1998); M. Baldo, A. Fiasconaro, H. Q. Song,
 G.~Giansiracusa, and U.~Lombardo, Phys. Rev. {\bf C65}, 017303 (2002).

\bibitem{wiringa:1995} R. B. Wiringa, V. G. J. Stoks, and R. Schialla,
Phys. Rev. {\bf C 51}, 38(1995).

\bibitem{baldo:1997} M.~Baldo, I.~Bombaci, and G.~F.Burgio,
Astron. Astrophys. {\bf 328}, 274 (1997).

\bibitem{grange:1989} P.~Grang\'e, A.~Lejeune, M.~Martzolff, and J.-F.Mathiot,
 Phys. Rev. {\bf C40}, 1040(1989).

\bibitem{machleidt:1989} R.~Machleidt, Adv. Nucl. Phys. {\bf 19}, 189 (1989).

\bibitem{wei:2002} W.~Zuo, A.~Lejeune, U.~Lombardo, and J.~F.~Mathiot,
Nucl.~Phys. {\bf A 706}, 418(2002); Eur. Phys. J. {\bf A 14}, 469(2002).

\bibitem{fuchs:2003} C. Fuchs, Lect. Notes Phys. {\bf 641}, 119 (2004).

\bibitem{lejeune:1986} A.~Lejeune, P.~Grange, M.~Martzolff, and J.~Cugnon,
Nucl.~Phys. {\bf A453}, 189 (1986).

\bibitem{walecka:1974} J. D. Walecka,
Ann. Phys. (N.Y.) {\bf 83}, (1974) 491; B. D. Serot and J. D.
Walecka, Adv. Nucl. Phys. {\bf 16}, (1986) 1;
 Int. Journ. Mod. Phys. {\bf E 6}, (1997) 515;
G. E. Brown, W. Weise, G. Baym and J. Speth,
 Comments Nucl. Part. Phys. {\bf 17}, (1987) 39.

\bibitem{baldo:1998} {\it Nuclear Methods and the Nuclear Equation of
State}, Ed. M. Baldo (World Scientific, Singapore, 1998), Chap.1.

\bibitem{donoghue:1985bu}
J.~F.~Donoghue and C.~R.~Nappi, Phys. Lett. {\bf B 168}, 105(1986).

\bibitem{dong:1996} S. J. Dong, J.-F. Laga\"{e}, K. F. Liu,
Phys. Rev. {\bf D 54}, (1996) 5496.

\bibitem{baym} G. Baym, Phys. Rev. Lett. {\bf 30}, 1340 (1973).

\bibitem{huber:1998} H.~Huber, F.~Weber and M.~K.~Weigel, {\it Phys.~Rev.}
{\bf C57}, 3484 (1998); Z. Y. Ma and L. Liu, Phys. Rev. {\bf C 66}, 024321 (2002).

\bibitem{greco:2001} V. Greco, M. Colonna, M. Di Toro, G. Fabbri, and F. Matera,
Phys. Rev. {\bf C 64}, 045203 (2001).

\bibitem{zhou:2004} X. R. Zhou, G. F. Burgio, U. Lombardo, H.-J. Schulze, and
W. Zuo, Phys. Rev. {\bf 69}, 018801 (2004).

\bibitem{baldo:1998prc} M.~Baldo, G.~F.Burgio, and H.-J. Schulze,
Phys. Rev. {\bf C 58}, 3688 (1998); {\bf C61}, 055801 (2000).

\end{thebibliography}
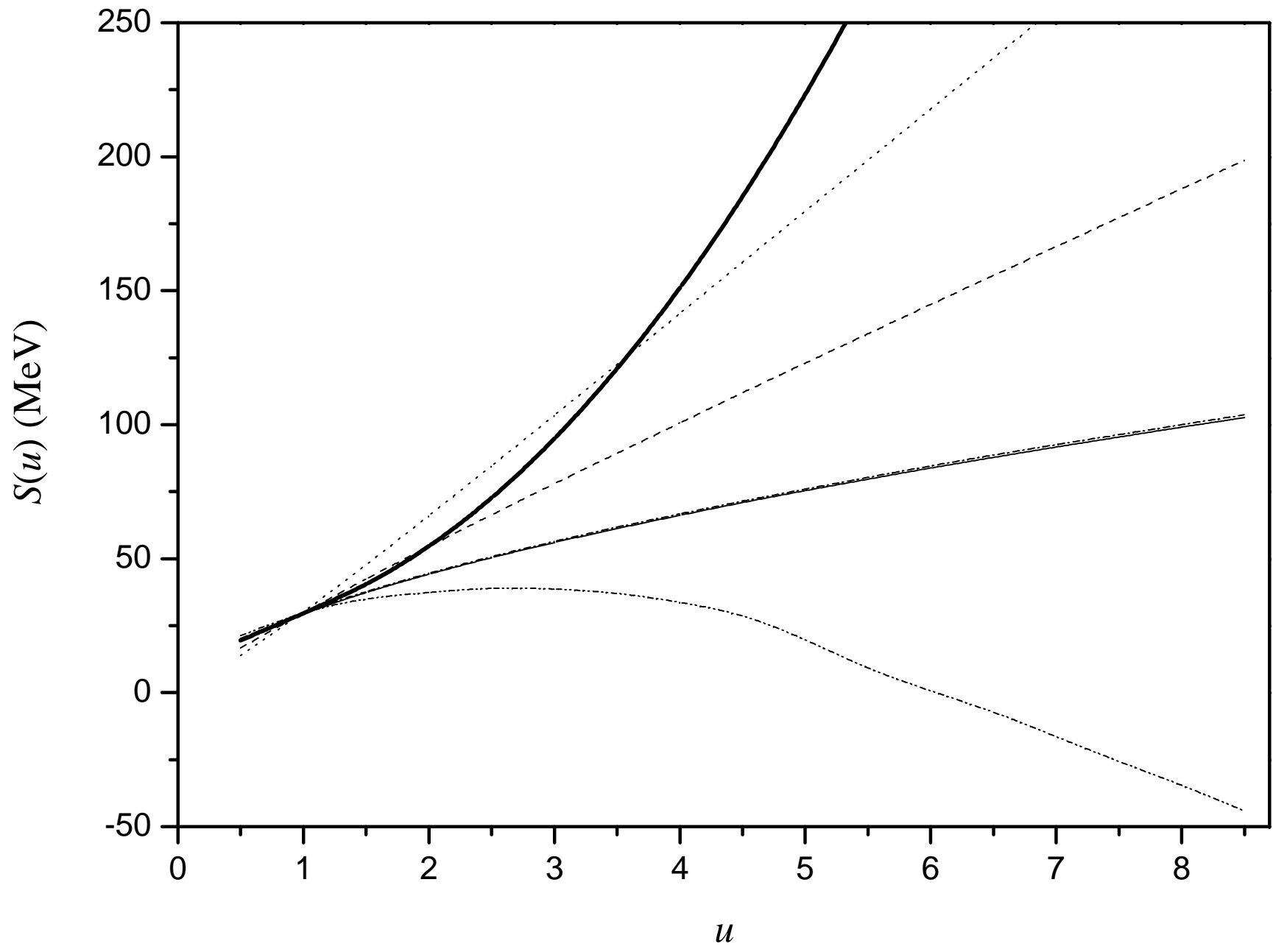
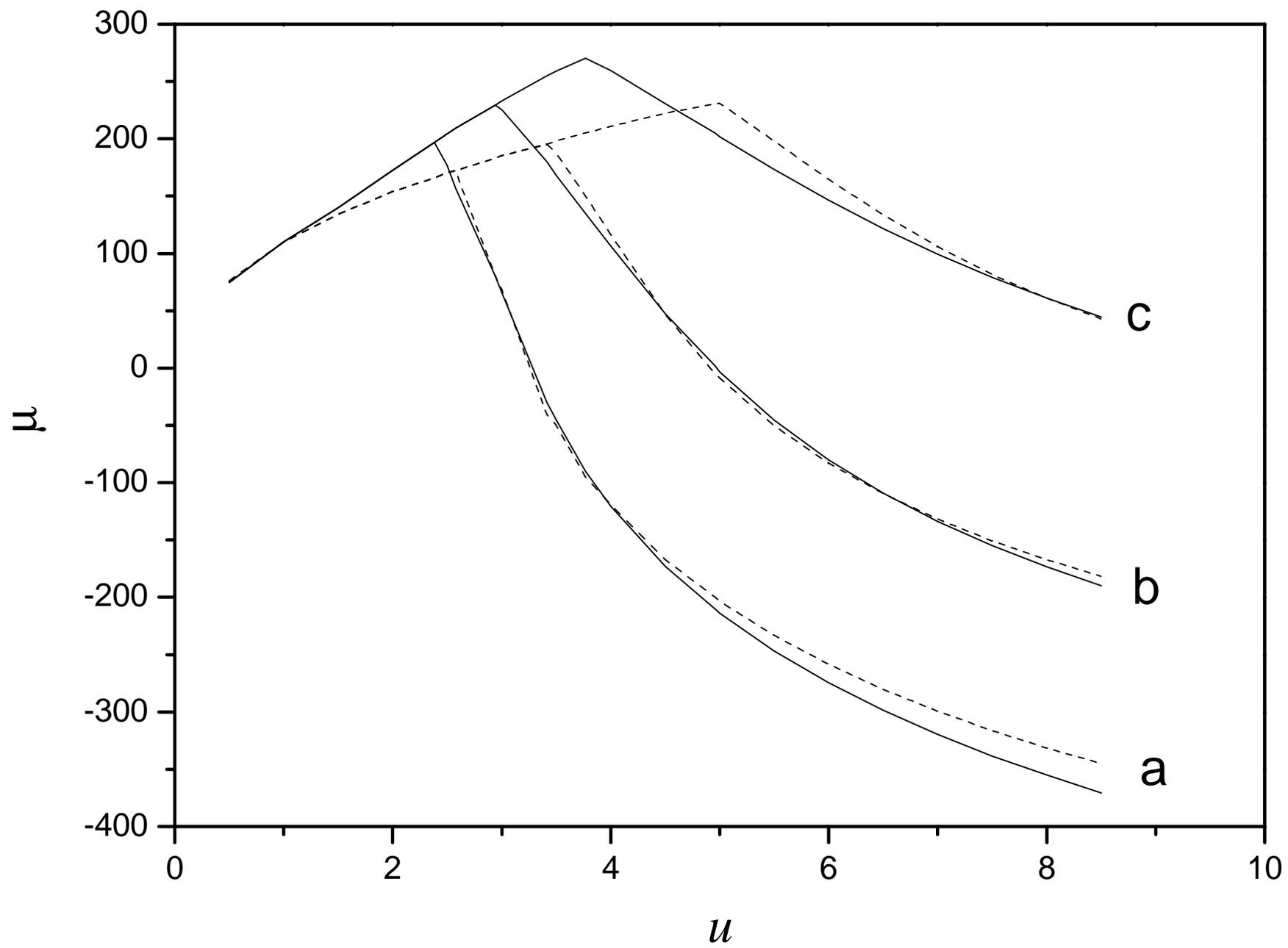
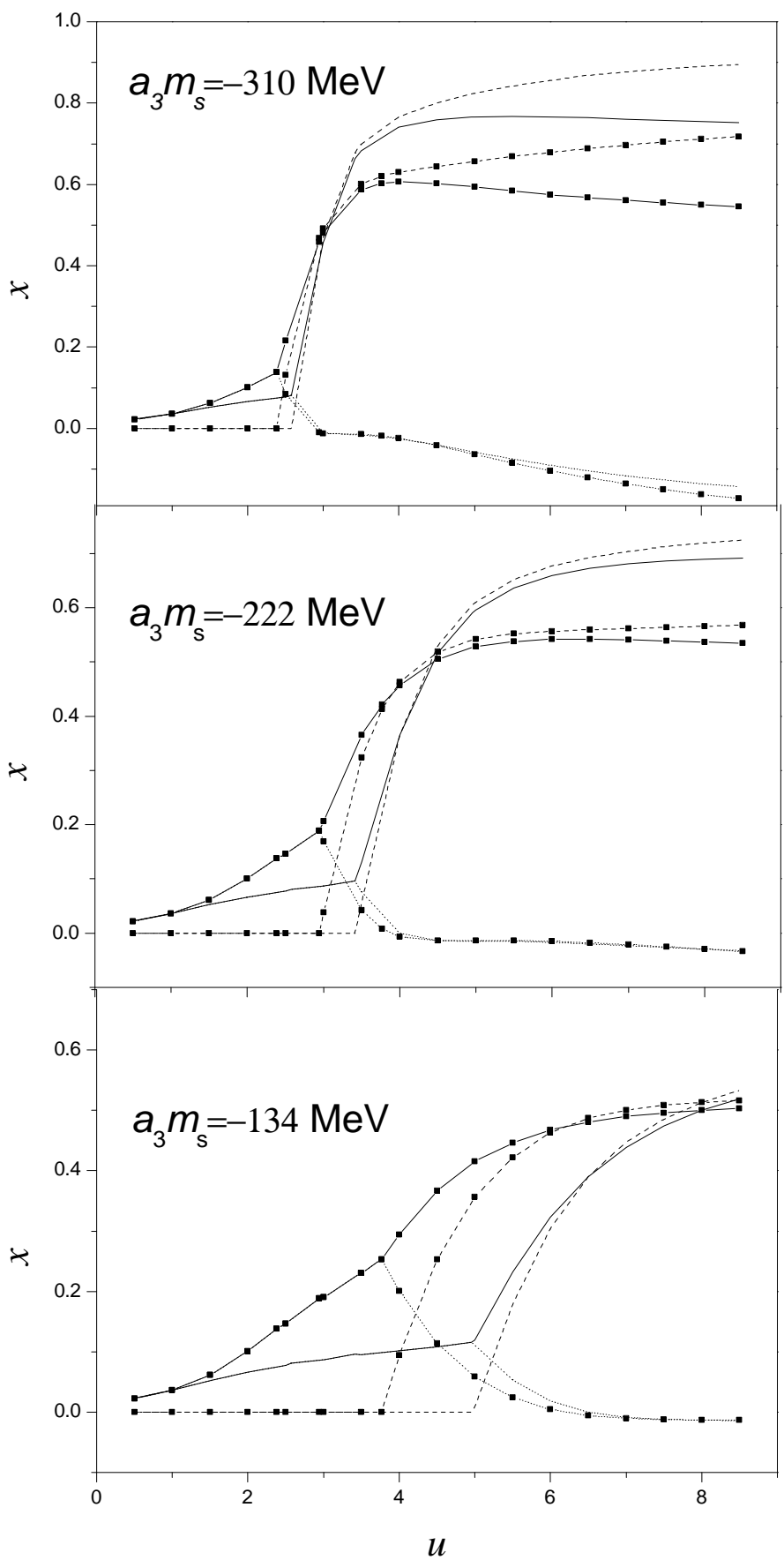
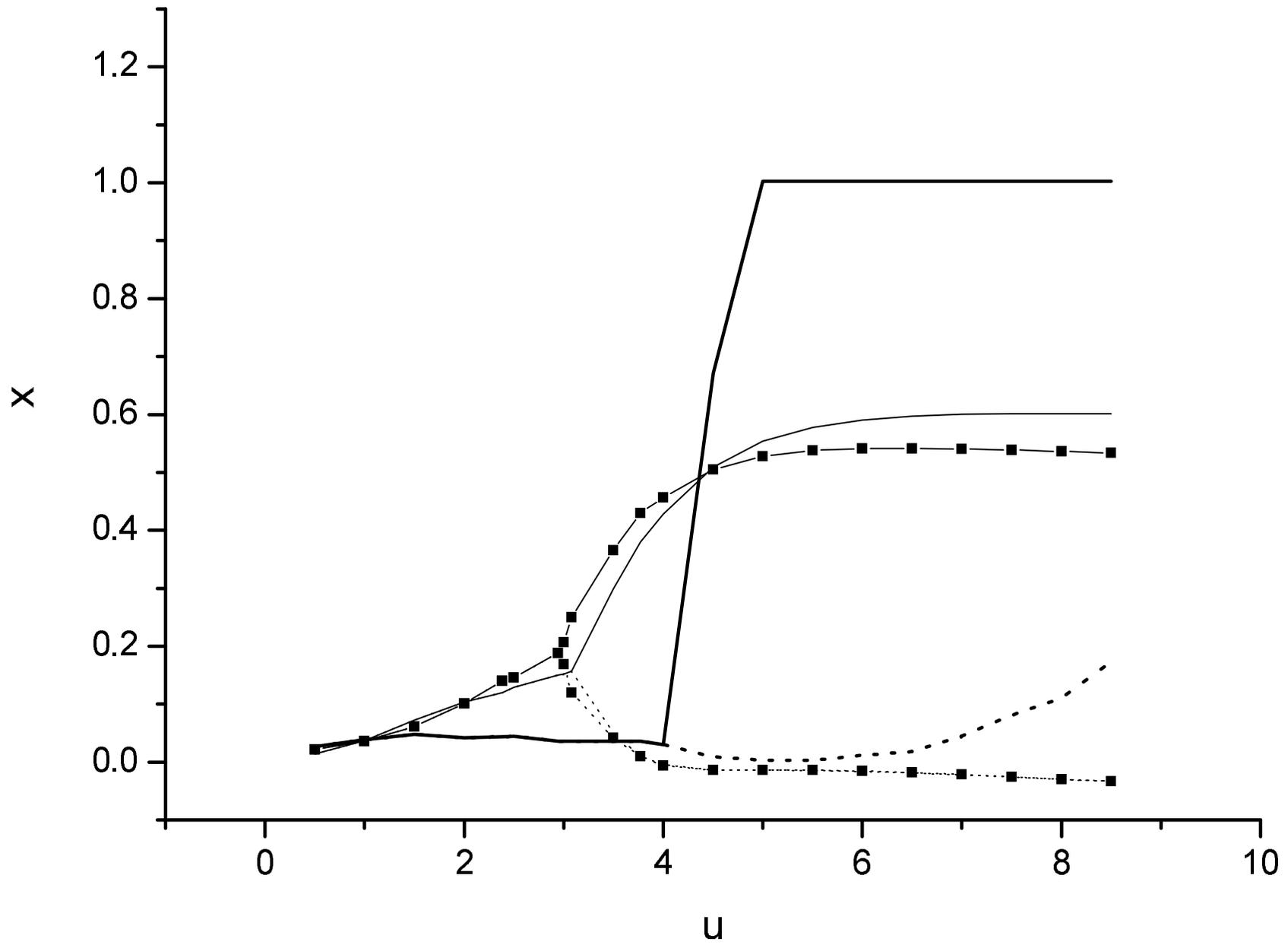
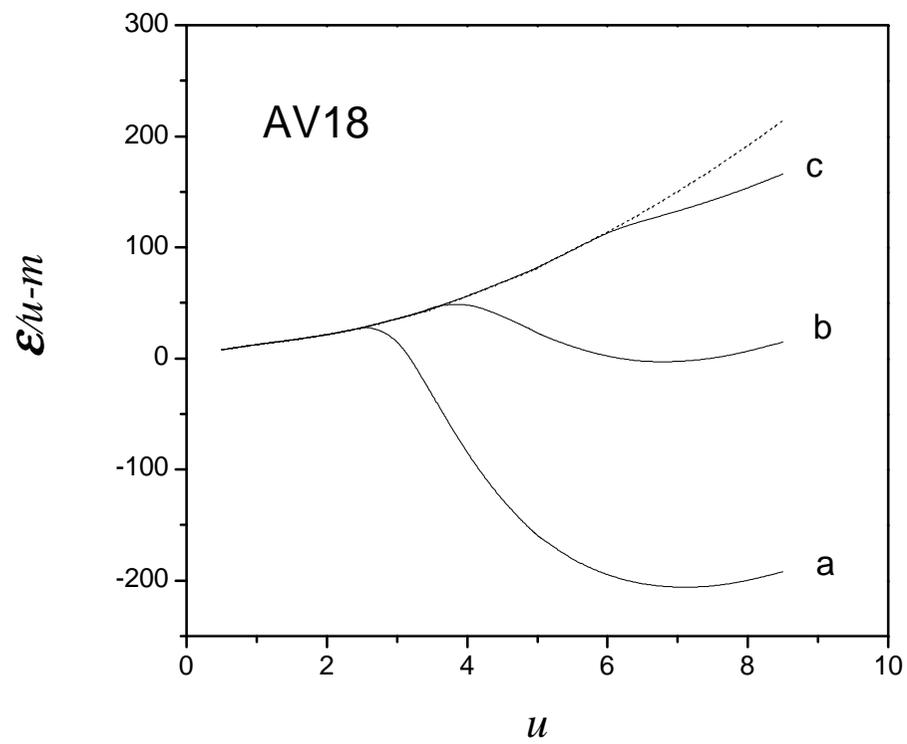
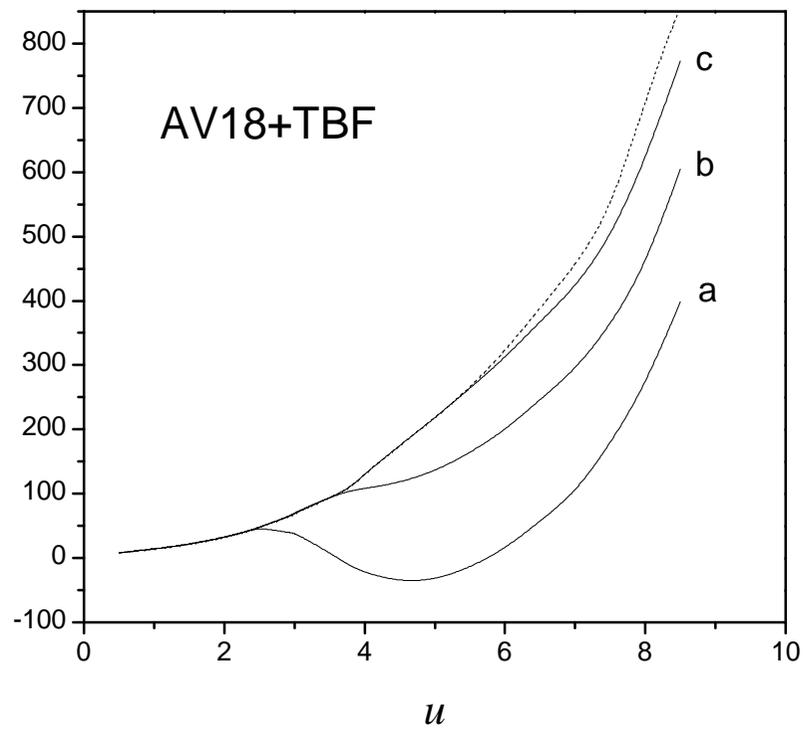
\end{document}